\documentclass{mn2e}
\usepackage{epsfig}
\usepackage{times}
\usepackage{hyperref}
\usepackage{graphicx}
\usepackage{amsfonts}
\usepackage{amssymb}
\usepackage{soul}
\usepackage{array}
\newcolumntype{C}[1]{>{\centering\arraybackslash}p{#1}}

\newcommand{\hi}{{\sc H\,i} }
\newcommand{\mnras}{MNRAS}
\newcommand{\nat}{Nature}
\newcommand{\aj}{AJ}
\newcommand{\apj}{ApJ}
\newcommand{\apjl}{ApJL}
\newcommand{\aap}{A\&A}

\newcommand{\araa}{ARA\&A}
\newcommand{\pasj}{PASJ}

\newcommand{\pasa}{PASA}
\newcommand{\physrep}{Physics Reports}
\newcommand{\nar}{New Astronomy Reviews}

\begin{document}

\title[Strongly lensed HI] 
{Strongly lensed neutral hydrogen emission: detection predictions with current and future radio interferometers}

\author[Deane, Obreschkow \& Heywood]
    {R.P. Deane$^{1,2}$\thanks{email: r.deane@ru.ac.za}, D. Obreschkow$^3$, I.~Heywood$^{4,1}$\\
 $^1$ Centre for Radio Astronomy Techniques and Technologies, Department of Physics and Electronics, Rhodes University, Grahamstown, South Africa \\
$^2$ Square Kilometre Array South Africa, Pinelands, Cape Town, South Africa\\
$^3$ International Centre for Radio Astronomy Research (ICRAR), M468, University of Western Australia, 35 Stirling Hwy, Crawley, WA 6009, Australia\\
$^4$ CSIRO Astronomy and Space Science, Marsfield, NSW, Australia
\\}
\maketitle
\begin{abstract}
Strong gravitational lensing provides some of the deepest views of the Universe, enabling studies of high-redshift galaxies only possible with next-generation facilities without the lensing phenomenon. To date, 21 cm radio emission from neutral hydrogen has only been detected directly out to $z\sim0.2$, limited by the sensitivity and instantaneous bandwidth of current radio telescopes. We discuss how current and future radio interferometers such as the Square Kilometre Array (SKA) will detect lensed \hi emission in individual galaxies at high redshift. Our calculations rely on a semi-analytic galaxy simulation with realistic \hi disks (by size, density profile and rotation), in a cosmological context, combined with general relativistic ray tracing. Wide-field, blind \hi surveys with the SKA are predicted to be efficient at discovering lensed \hi systems, increasingly so at $z\gtrsim2$. This will be enabled by the combination of the magnification boosts, the steepness of the \hi luminosity function at the high-mass end, and the fact that the \hi spectral line is relatively isolated in frequency. These surveys will simultaneously provide a new technique for foreground lens selection and yield the highest redshift \hi emission detections. More near term (and existing) cm-wave facilities will push the high redshift \hi envelope through targeted surveys of known lenses.
\end{abstract}

\section{Introduction}

The evolution of neutral cold gas ($T \lesssim10^4$~K) is fundamental to our understanding of galaxy evolution. Our current cosmological picture is that following recombination at $z\sim1100$, the Universe was predominately neutral until reionisation occurred between $z\sim6-10$, driven (in poorly constrained proportion) by the onset of star formation and supermassive black hole accretion in the earliest galactic haloes. The intergalactic medium is rapidly ionized about 1~Gyr after the Big Bang, with the vast majority of remaining neutral hydrogen found in galaxies where it is sufficiently shielded and/or replenished (e.g. Barkana \& Loeb, 2001; Fan et al.~2006).

Within galactic haloes, cold gas is a significant mass component and evolutionary ingredient: it is the pristine fuel for star formation, as well as the exhaust of supernova explosions. The biggest obstacle in uncovering the precise evolutionary role of \hi is that direct H{\sc\,i} observations are currently restricted to the modern Universe ($z\lesssim0.2$). Many important questions remain observationally unanswered as a result: does the cold gas continually condense onto galactic disks, resupplying that which is driven out by AGN and supernovae (e.g. Dav{\'e} et al.~2013, Morganti et al.~2013)? How do H{\sc\,i} morphology and dispersion evolve out to cosmic noon ($z\sim2$) and beyond? What is the link between the evolution of H{\sc\,i} and cosmic star formation history of the Universe (e.g. Obreschkow \& Rawlings 2009)? Addressing these questions demands H{\sc\,i} observations at look-back times much greater than $\sim\!10$ dynamical times (i.e., at $z\gtrsim0.2$). Since H{\sc\,i} extends to large galactocentric radii, \mbox{high-$z$} H{\sc\,i} observations may also enable studies of the halo structure evolution and allow an important measurement of the baryon angular momentum (Obreschkow~et~al.~2015). 

Over the next decade we are set to revolutionize our view of the cosmic evolution of neutral hydrogen. A broad range of radio telescopes are being developed to tune into the rest-frame 21~cm spectral line out to large redshifts. There are a number of upcoming radio facilities, including the Australian Square Kilometre Array Pathfinder (ASKAP; Johnston et al., 2007), MeerKAT (Jonas, 2009), SKA (e.g. Dewdney et al.~2013), as well the upgraded Karl G. Jansky Very Large Array (VLA), Westerbork Synthesis Radio Telescope (WSRT) with the APERture Tile In Focus project (APERTIF, Verheijen et al.~2008), and the Giant Metrewave Radio Telescope (GMRT, Ananthakrishnan, 2005). These telescopes will all perform deep \hi~surveys, some of which will detect individual galaxies out to redshifts of $z\sim1$ and beyond. These telescopes will enable the study of \hi~in galaxies with sample sizes and at distances not practically possible with current facilities. This will renew our understanding of neutral hydrogen and its role in galaxy evolution. While \hi~absorption studies have made important contributions to this field (e.g. Gupta 2006, Allison et al.~2015), there remain significant sources of uncertainty in absorption systems that limit extrapolation to the cosmic evolution of \hi~in galaxies (e.g. spin temperature, gas morphology). 

One of the byproducts of the significantly larger survey speed and extended redshift range of next-generation instruments will be the dramatically increased probability that an {\sc H\,i}-detected source is strongly lensed by a foreground galaxy. Here we use the canonical definition of strongly lensed as a total magnification $\mu\geq2$. This corresponds to a factor $\mu^2\geq4$ saving in integration time to detect a given source, provided it is unresolved in the image plane. This will enable the detection of \hi at even greater cosmological distances than otherwise expected with future instruments.

In this letter, we use an N-body simulation with semi-analytic prescriptions for the spatially-extended \hi~sources to predict the magnification statistics and resultant number counts of strongly lensed \hi emission. Our simulation uses a 150~deg$^2$ field of view mock observing cone out to a redshift limit of $z_{\rm source}<4$, enabling the prediction of detection rates of future \hi surveys. We assume cosmological parameter values of $\Omega_{\rm M}$ = 0.25, $\Omega_{\Lambda}$ = 0.75, and $H_0$ = 73 km\,s$^{-1}$\,Mpc$^{-1}$ (Millennium run cosmology, Springel et al. 2005). The best estimate of the cosmological parameters today (e.g. the latest Planck cosmological parameters, Planck collaboration~2015) are slightly different, but this has a marginal difference on both the semi-analytic results (Guo et al., 2013) and magnification estimates, and thus on our predictions.

\section{Simulation}\label{sec:sim}

\subsection{Mock sky}

The traditional approach in estimating the lensing probability (also known as the lensing opacity) is to use analytic volume densities of both the foreground and background source populations and integrate for a given redshift range. This approach has been readily applied in lensing predictions for point-like sources (e.g. optical quasars, high redshift supernovae, Oguri \& Marshall 2010; compact radio AGN, McKean et al.~2015), however, the large spatial extent and non-trivial density-velocity structure of \hi disks pose a challenge to estimating total magnification for \hi sources. This challenge is further complicated by the requirement of modelling large cosmological volumes ($\gg 10^6$ galaxies) in order to obtain sufficiently converged statistics to support actual galaxy surveys. We address this challenge using the S$^3$-SAX model (Obreschkow et al.~2009), a semi-analytic model (SAM) extending the SAMs by Croton et al. (2006) and De Lucia \& Blaizot (2007), relying on the Millennium $N$-body simulation of cosmic structure (Springel et al., 2005). S$^3$-SAX is the only model to date that provides a mock sky of millions of \hi disks with realistic spatial distributions and velocity profiles (Obreschkow et al., 2009b). This model is well-suited to generate \hi lensing predictions since the mock galaxies have all the properties needed for both the source and lens galaxies. Explicitly, (i) the mock galaxies are correctly clustered due to the underlying Millennium simulation; (ii) the \hi (needed in source-galaxies) has realistic spatial distributions, accounting for the large \hi disk size ($>$ stellar effective radius) and central troughs; (iii) the \hi has realistic velocity profiles that also satisfy the Tully-Fisher relation and local \hi velocity function (Obreschkow et al. 2013); (iv) the galaxies have multiple components (halo+disk+bulge) with realistic mass distributions (needed for lens galaxies), (v) the properties of the mock galaxies evolve with redshift according to the prescriptions of Croton et al. (2006) and Obreschkow et al.~(2009). A velocity dispersion of $\sigma_{\rm v,HI} = 10$~km\,s$^{-1}$ is added to the SAM-generated \hi velocity profile, consistent with observed properties of nearby \hi galaxies (e.g. de Blok et al.~2008). The typical HI velocity dispersions and scale heights potentially increase with redshift, similarly to the observed increase in the molecular velocity dispersion (e.g. F{\"o}rster Schreiber et al. 2006). We decided to neglect this effect, since the rest-frame channel widths and pixel sizes of our lensing calculations are generally a few times larger than the dispersion scale and thickness of the disk. Crucially, the use of an N-body simulation and full source-to-lens plane ray-tracing enables the \hi flux of all lensed galaxies to be split into individual frequency channels, resulting in a realistic prediction of the observables (e.g. peak channel flux, velocity line widths) and hence optimal detection strategies.

 \begin{figure}
\centerline{\epsfig{file=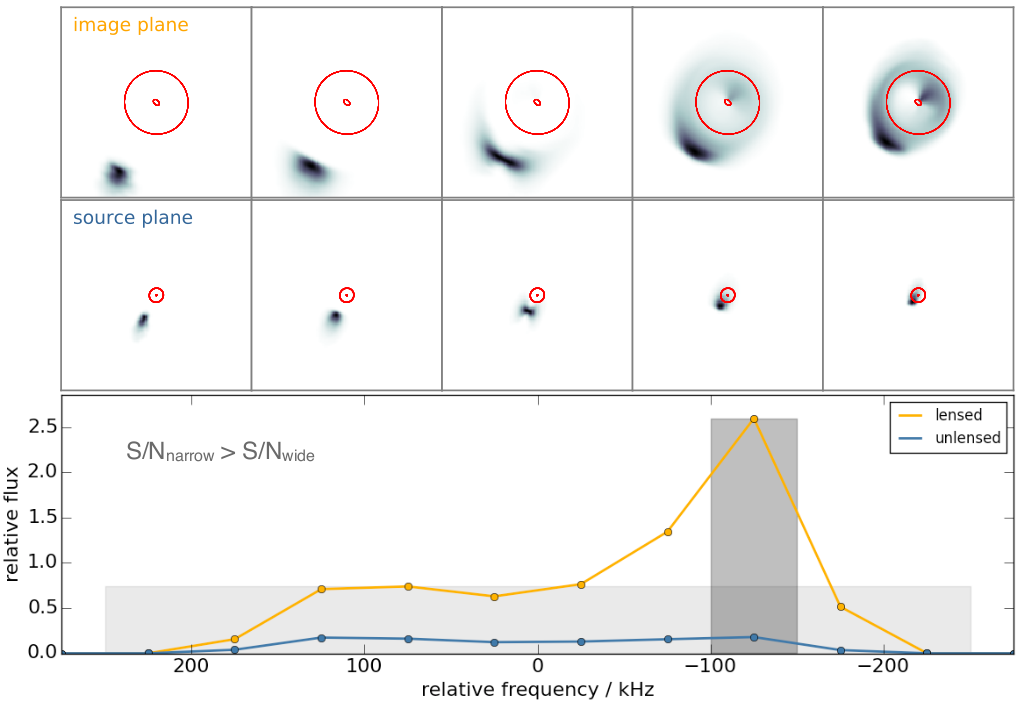, angle=0, width=\columnwidth}}
\caption{{\bf Top: }Image and source plane channel maps of a $z = 1.86$, $M_{\rm HI} = 2.7 \times 10^9$~M$_\odot$ \hi disk (frequency resolution = 200 kHz), demonstrating how the approaching \hi emission overlaps with the inner lens caustic and results in a full \hi Einstein ring in some channels. The lens is at a redshift of $z_{\rm l} = 0.58$, with a total mass $M_{\rm halo} = 3.6 \times 10^{13}$~M$_\odot$ and bulge ellipticity $e_{\rm b} = 0.24$.  Each frame is 6.4$\times$6.4~arcsec$^2$. The caustics/critical curves are plotted in red. {\bf Bottom: }Lensed and unlensed \hi profiles of the same system at higher frequency resolution (50 kHz). Significant differential magnification is seen between channels to the extent that the probability of detection is higher for a narrow ($\sim$50 kHz) channel width, rather than the full \hi line width. This argues that the detection statistics presented in this paper are likely to be conservative if source detection algorithms are also run on narrow channel widths. } 
\label{fig:chanmaps}
\end{figure}

\subsection{Lens candidate selection}\label{subsection_selection}

 Based on the S$^3$-SAX model we produce five statistically independent mock observing cones (as described in Obreschkow et al., 2009b), each covering a field-of-view of 30 deg$^2$ and extending from $z=0$ to $z=4$ (30 deg$^2$ corresponds to the cross-section of the Millennium box at $z\approx4$). Together these five cones represent a mock sky covering 150 deg$^2$ with negligible cosmic variance. From this sky we extract all lens-source pair candidates with a background source integrated \hi flux $S^{\rm int}_\mathrm{\rm HI,s} > 100 \, \mu$Jy\,km\,s$^{-1}$ and foreground halo virial mass $M_\mathrm{\rm vir,l} > 5\times10^{11}\,\mathrm{M}_\odot$. In addition, we only consider lens candidates with a source redshift $z_{\rm source} < 4$ and an intrinsic lens-source impact factor $< (20$~arcsec) AND ($0.15\,R_{\rm vir}$), where $R_{\rm vir}$ is the foreground halo virial radius. These cuts dramatically decrease the computational demand and have a negligible effect on the inferred lensing statistics (following a trial and error process). For example, the majority of lensing galaxies have Einstein radii (and hence deflection angles) of order a few arcseconds (e.g. Treu 2010, and references therein). The above selection criteria result in $\sim10^6$ lens-source pair candidates, i.e. approximately 2 per arcmin$^2$.

\subsection{\hi magnification factor}\label{subsection_mu}

The lens models are split into three components: (a) a dark matter halo, (b) bulge, and (c) a disk component. The dark matter halo is assumed to have a spherically-symmetric NFW density profile with a concentration parameter derived from a redshift and halo mass dependent fit (Hennawi et al.~2007), the distribution of which is consistent with empirical results from X-ray and strong lensing observations of galaxy clusters (Comerford \& Natarajan 2007). The latter consistency is particularly important since magnification bias will preferentially select high surface mass density foreground dark matter haloes.

The bulge and disk effective radii are derived directly from the self-consistent dynamics of the semi-analytic prescription to all galaxies in the simulation. The bulge is set to a de Vaucoleurs density profile with an ellipticity drawn from a Gaussian distribution with mean and 1$\sigma$ standard deviation of $\mu_{\rm e} = 0.3 \pm 0.3$ (truncated between $0 < \mu_{\rm e} < 0.98$), consistent with observational properties of bulges and previous approaches (Jorgensen \& Franx 1994, M\"oller et al. 2007, Oguri \& Marshall 2010). The halo+bulge density profiles used in SAX-Lens are realistic not only in that the distribution of their concentration parameters match available observations; but also in that global \hi linewidths (outer rotation) and CO linewidths (inner rotation) are both consistent with observations (Obreschkow et al. 2009b). The disk is modeled as a spherically-symmetric component with an exponential density profile. It is sub-dominant at all radii and adds a marginal fraction of total convergence to the system as seen in M\"oller et al. (2007). We do not consider the additional convergence contributed by satellite or neighbouring galaxies associated with the foreground lens, which may add up to several tens of percent to some of the largest image separations (and hence magnification, Oguri 2006), however this effect decreases for lower mass, lower redshift lenses as discussed by M\"oller et al. (2007).

 With both source and lens profiles defined, we perform the general-relativistic ray-tracing using the {\tt glafic} software package (Oguri 2010). To this end, the \hi emission of all sources is split into 50 kHz channels (intrinsic velocity widths of $\sim11-53$~km\,s$^{-1}$ for $z = 0-4$), each of which is individually lensed. We thus have noise-free image plane (lensed) and source plane (unlensed) observing cubes (two spatial dimensions, one frequency dimension) of all \hi sources. Fig.~\ref{fig:chanmaps} shows example source and image plane channel maps of a $z\sim2$ source (at 200~kHz channel resolution for clarity). The associated lensed and unlensed global profiles are shown in the bottom panel. Optimal detection strategies for such a distorted profile are discussed in Sec.~\ref{sec:distort}. The global magnification factor $\mu$ is defined as the ratio between the integrated \hi fluxes of the magnified image and the unmagnified source. All systems with $\mu\geq2$ are written to a catalogue, referred to as the Semi-Analytic eXtragalalactic Lens (SAX-Lens) catalogue. This catalog contains $45,734$ lens-source pairs, i.e.\ about 300 per deg$^2$.

\section{Predictions}

\subsection{Survey-independent lens statistics}

\begin{figure}
\centerline{\epsfig{file=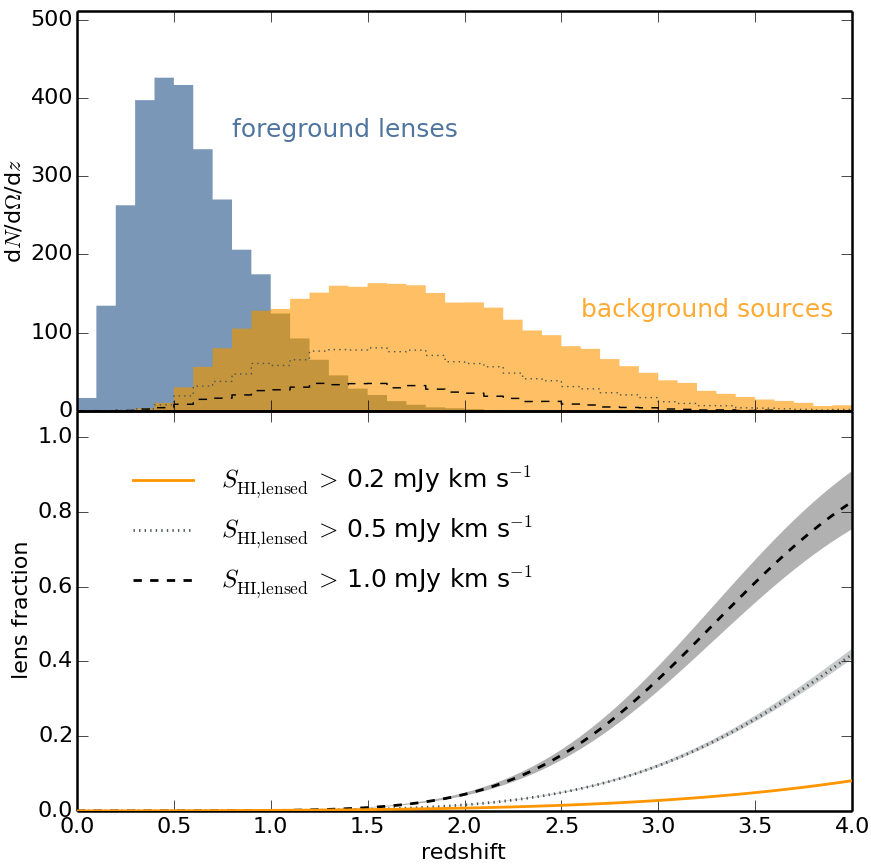, angle=0, width=\columnwidth}}
\caption{{\bf Top: }Redshift distribution of all foreground lens and background source pairs with total \hi magnification $\mu > 2$ ($\Omega$ units: deg$^{-2}$). {\bf Bottom: }The fraction of \hi sources that are predicted to be strongly lensed (lensing opacity), given a velocity-integrated flux threshold (see legend). The grey shadings indicate the 1$\sigma$ Poisson error, given the full 150 deg$^2$ simulation. This forecasts the efficiency of future \hi surveys at selecting lensed systems, particularly if cross-correlated with massive foreground galaxies at lower redshift.}
\label{fig:zhist}
\end{figure}

The primary output of the methodology laid out in Sec.~\ref{sec:sim} is the SAX-Lens catalogue of simulated \hi sources that are magnified by a factor of 2 or more. In Fig.~\ref{fig:zhist} (top panel) we show the redshift distribution of both the foreground lenses and the background \hi sources. This demonstrates the lensing probability only becomes appreciable above $z \gtrsim 0.5$, well above the highest redshift \hi emission detection to date ($z\sim0.2$, Catinella et al.~2008). The distribution of strongly lensed \hi disks with $S_\mathrm{HI,s} > 100 \, \mu$Jy\,km\,s$^{-1}$ peaks at $z\sim 1.5-1.8$, which corresponds to an observed frequency of $\nu_{\rm obs} \sim 570-510$~MHz. The bottom panel in Fig.~\ref{fig:zhist} demonstrates that higher redshift sources are, on average, more magnified due to higher alignment probability and smaller apparent source size. This means that surveys designed to detect these sources will be highly biased towards strongly lensed systems. It is likely that future radio telescopes able to detect lensed \hi sources, will, due to the wide ($\geq2:1$) instantaneous band widths of modern receivers, simultaneously detect the \hi emission or absorption of the foreground lens. This will greatly expedite the followup observations of lensed sources discovered with an entirely different selection function to other lens catalogues (e.g. CLASS, Myers et al.~2003; SLACS, Bolton~et~al.~2006).

\begin{figure}
\centerline{\epsfig{file=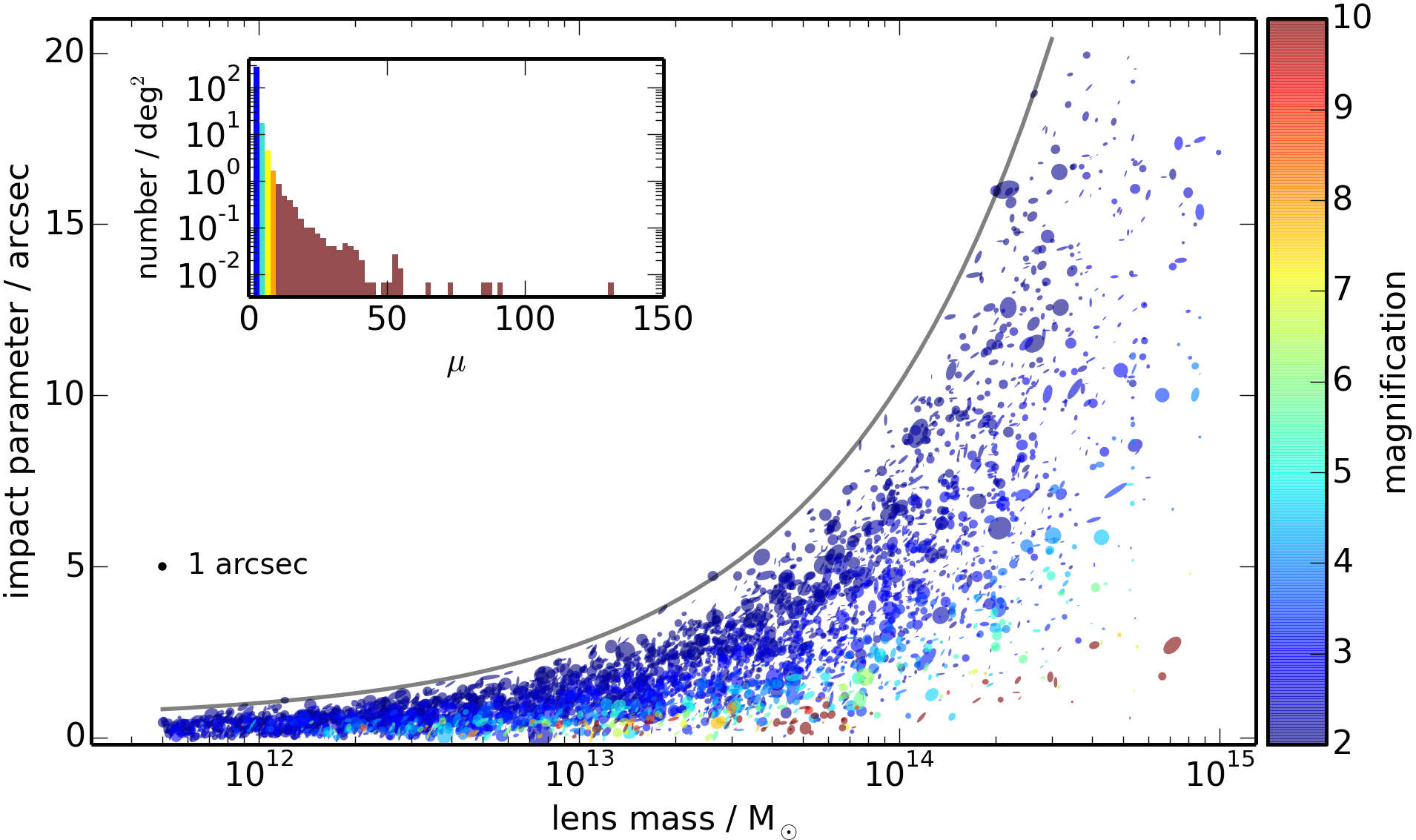, angle=0, width=\columnwidth}}
\caption{Total magnification as a function of foreground lens virial mass and impact parameter. The magnification is clipped to $\mu = 10$ and only every 10$^{\rm th}$ lensed source in the full 150~deg$^2$ simulation is shown. Ellipses show the intrinsic angular size of the background \hi disks, with a 1~arcsec reference shown in black (below inset). This shows that the highest magnification sources tend to have small angular size. The ellipse position angles represent the angle between the \hi semi-major axis and the impact vector. Vertical (horizontal) semi-major axes represent aligned (perpendicular) impact vectors and \hi semi-major axes. The dark grey line shows the fit to the upper envelope (see text). {\sl Inset:} Total magnification distribution of $\mu\geq2$ sources based on full 150 deg$^2$ simulation (bin width = 2).} 
\label{fig:LMhalo}
\end{figure}

In Fig.~\ref{fig:LMhalo} we show some of the statistical properties of the lens-source pairs, including the foreground lens halo (total) mass, impact parameter (angular distance between lens and source centroids), magnification (colour scale), as well as the \hi disk angular size. The ellipse axis ratios correspond to that of the background \hi source (with min=0.01). The ellipse position angle represents the angle between the impact vector and \hi semi-major axis (see caption). Fig.~\ref{fig:LMhalo} shows a rough envelope above which virtually no sources are strongly lensed (marginalized over a number of parameters, e.g. redshifts, halo concentration parameter). An approximate fit to this envelope is: impact~$\sim 0.50 + (M_{\rm lens}/10^{12.88}~{\rm M}_\odot)^{0.65}$ arcsec.

The highest magnification sources ($\mu \gtrsim 10$) in the SAX-Lens catalogue have an Einstein ring (see Fig.~\ref{fig:chanmaps}) in at least one of their channel maps and thus have the largest apparent solid angles. However, current and future radio facilities will not have sufficient angular resolution (for the vast majority of their collecting area) at the appropriate frequency to spatially resolve the \hi Einstein rings. This maximises their probability of detection since the surveys will benefit from the total \hi magnification boost. This also implies that macro lens models derived from optical/NIR/mm observations will be required to constrain the intrinsic \hi disk mass and size. Of all the lensed \hi disks in the SAX-Lens catalogue, just 6 percent have an extent larger than 0.5~arcsec. This implies that even SKA1-MID, with a $< 0.5$~arcsec PSF for $\nu_{\rm obs} \lesssim 1$~GHz, will not resolve the vast majority of lensed \hi disks. Of course, the sources that are resolved will be of great interest for individual followup to perform more detailed modelling of the astrophysical properties. In this paper we limit the scope to the detection statistics, assuming all sources are unresolved.

\subsection{Survey-specific predictions}

\begin{table}
\begin{tabular}{lC{0.6cm}C{0.7cm}ccC{0.7cm}}
\hline
{\bf Survey} &   $\Omega$ &  $\sigma_{\rm 100kHz}$  &   $z_{\rm max}$  &   $N_{\rm detect}$ &  $\bar{z}_{\rm detect}$  \\ 
                    &   deg$^2$    &             $\mu$Jy/b                    &       &     &   \\ 
 \hline \hline 
CHILES$^a$   			&  0.25 	& 20 	& 0.45 	& $<0.1$ 		& 0.35  \\
LADUMA$^b$   		&  0.60 	& 8 	& 1.45 	& 20 (47) 		& 0.96  \\
DINGO-UD$^c$   		&  60 	& 38 	& 0.43 	& 3 (8) 		& 0.30  \\
DINGO-D$^c$   		&  150 	& 85 	& 0.26 	& 1 (2)		& 0.16  \\
SKA1-Deep   		&  100 	& 10 	& 3.06 	& 2,880 (7,265) 	& 1.26  \\
SKA1-MedDeep   	&  1,000 	& 33 	& 3.06 	& 2,480 (7,193)	& 1.19  \\
SKA1-Wide   		&  10,000	&100& 3.06 	& 2,667 (7,467)	& 0.93  \\
\hline
\end{tabular}
   \caption{Predictions for the number of {\sc Hi}-detectable lensed systems in future surveys. As described in the main text, a detection is defined as $S_{\rm mean} > 5\sigma$. For the single pointing surveys (CHILES, LADUMA), the area ($\Omega$) corresponds to the 1.4 GHz field of view, which scales as ($1 + z$)$^2$. SKA surveys assume the SKA1-MID specifications, $\sim$1 year integrations and indicative survey areas. However, we note that SKA1-LOW could cover the $z \gtrsim 2$ space if its upper frequency limit were to be increased to $\sim$470~MHz. The SKA1-MID specifications are drawn from Dewdney et al. (2013), but the total number of antennas decreased from 254 to 197 in line with the 2015 re-baselining process. The quoted SKA1-MID survey sensitivities are assumed for 1420.4 MHz and 1050 MHz, where the latter is the highest frequency in Band 1. Sensitivities for both SKA Band 1 and 2 scale as $(1 + z)^{-1}$ due to the increase field-of-view but fixed survey area. Multi-wavelength cross-correlation may reveal many more \hi lenses, as suggested by the 3$\sigma$ predictions in parentheses. Note that the APERTIF surveys resulted in zero detections in our simulation, so detection probability estimates were not possible. For reference, the approximate number of strongly lensed systems known at present is $\sim 500$. \newline
{  \scriptsize
  $^a$ COSMOS HI Large Extragalactic Survey (Fernandez et al., 2015) \newline
  $^b$ Looking at the Distant Universe with the MeerKAT Array (Holwerda, Blyth \& Baker, 2012) \newline
  $^c$ Deep Investigations of Neutral Gas Origins (Meyer, 2009)} }
   \label{tab:surveys}
\end{table}

Having presented the basic statistical properties of the lensed \hi systems, we now turn to predicted detection rates with planned surveys. There are a wide range of \hi surveys planned over the course of the next decade which will dramatically expand our view and understanding of \hi in galaxies. In Tab.~\ref{tab:surveys} we list the predicted number of \hi lenses (and mean redshift) that will be detectable for some of the major future surveys. We define the detection threshold as $\bar{S}_{\rm HI} > 5\sigma$, where $\bar{S}_{\rm HI}$ the mean \hi profile flux in units of Jy and the noise corresponds to that expected in a channel width equal to the \hi profile FWHM for each individual galaxy. The latter are determined from the 100 kHz channel sensitivities listed in Tab.~\ref{tab:surveys}. A major caveat in any such prediction is the ability to distinguish lensed from non-lensed sources, given that they will be unresolved (in \hi) and the vast majority will only be lensed by a factor of two (see inset, Fig.~\ref{fig:LMhalo}). Multi-wavelength data will therefore be essential to unambiguously identify low ($\mu \lesssim 4$) magnification \hi lenses, but may also be used to lower the detection threshold and discover many more lensed systems, as suggested by the number of $3\sigma$ detections indicated in parentheses.

Table~\ref{tab:surveys} shows that none of the pre-SKA surveys are particularly optimized for detecting large numbers of lensed \hi systems (e.g. $\sim4\--20$ detections for DINGO and LADUMA). However, lensed \hi systems will increase the number counts and enhance these already planned \hi surveys beyond their original science cases. The predicted mean redshifts of these lensed \hi sources are much higher than the current record of $z \sim 0.2$ (except for DINGO-D). This combined with the magnification boosts implies that it will be possible to detect high redshift systems with much lower \hi mass than that possible for non-lensed systems. Furthermore, the relative isolation of the \hi line, the steep high-end \hi mass function and the systematically larger magnification factors at higher redshift suggest that some of the most distant \hi sources will be detected and identified within a fraction of the total survey duration. See Serjeant (2014) for a discussion on using the high-mass end of the \hi mass function to select lenses. The above factors, combined with the increasing \hi lens opacity at higher redshift (see Fig.~\ref{fig:zhist}, bottom panel) will make a significant impact on $\Omega_{\rm HI}$ measurements at $z \gtrsim 2$. Indeed, even $z\sim1$ $\Omega_{\rm HI}$ values with pre-SKA facilities may require strong-lensing correction factors. Tab.~\ref{tab:surveys} also suggests that targeted surveys for lensed \hi may be the optimal strategy in the pre-SKA era. 

The SKA1 surveys will revolutionize our view of \hi in galaxies at high redshift. The predictions suggest $\sim0.8-2.2 \times 10^4$ lensed detections, well over an order of magnitude larger than the presently known lens population at any wavelength. The mean and maximum redshifts of  sources detected in these surveys will provide important insights on the evolution of cold gas towards cosmic noon ($z\sim2$), a perspective not possible with any other facility. 

The full SKA is expected to detect up to $\sim10^9$ non-lensed \hi galaxies out to $z \sim1$ through surveys aimed primarily at high precision measurements of baryonic acoustic oscillations (Rawlings~et~al.~2004, Maartens et al.~2015, Bull et al.~2015). However, even with the SKA, strong gravitational lensing will be required to characterize Milky Way like galaxies to the epoch of peak cosmic star formation and AGN activity at $z \sim 2-3$. Stacking methods have been proposed to pursue this goal, however, the uncertainties and associated contamination using these methods are unclear. Furthermore, calibration and source modelling limitations may place upper limits on the sensitivity of stacking methods (Makhathini et al., in prep.). Well-modeled strong gravitational lenses, with the benefit of excellent multi-wavelength coverage from ALMA, LSST, JWST and the E-ELT will clearly provide an excellent opportunity to push back this observational frontier. While these surveys will dramatically enhance our knowledge of high redshift cold gas and disk formation, a further valuable contribution will be a new probe of the foreground lens population using a unique selection function that is not biased by dust attenuation (in both the source and lens galaxies).

\subsection{Optimal detection strategy}\label{sec:distort}

In all of the lensed statistics presented above, the detection thresholds assume a channel width matched to the FWHM of the \hi line width. This optimizes the S/N ratio and hence maximizes detection probability for the majority of lensed systems. However, there are cases when a narrow channel width has a significantly larger magnification than that of the total intensity map. This is simply due to the channel flux solid angle and position relative to a caustic. Such differential magnification is seen in the relative magnification of the CO (1-0) channel maps of a $z\sim2.3$ lensed disk galaxy (Deane et al.~2013a), as well as the order of magnitude difference in magnification of the radio core to CO emission (Deane et al.~2013b). Differential magnification can be exploited to increase the detection probability by running spectral line source finders with narrow channel widths as well as those optimized to the full galaxy line width. An illustration of this effect is seen in Fig.~\ref{fig:chanmaps} which shows a $z\sim2$ \hi source with total magnification $\mu\sim3.3$, however one particular 50~kHz channel is magnified by a factor $\mu > 10$. An appropriate channel selection yields a higher probability of detection than matching to the line width. Clearly, there is a great deal of optimization to be explored in both targeted and blind lensed \hi surveys, which we investigate in an upcoming paper.

\section{Conclusions}

The use of an N-body simulation with applied semi-analytics enables the following predictions relevant to lensed {\sc Hi}.

\begin{enumerate}

\item A full ray-tracing analysis of $10^6$ $z<4$ lens candidates in the 5$\times$30 deg$^2$ simulation reveal $\sim$46,000 strong lenses ($\mu>2$) with unlensed velocity-integrated \hi fluxes $S_{\rm HI} > 100~\mu$Jy\,km\,s$^{-1}$.

\item The predicted number counts imply that the dramatic increase in sensitivity and frequency coverage of upcoming radio facilities will enable \hi detection at redshifts where strong lensing has been highly beneficial to galaxy evolution studies that utilise other spectral lines (e.g. CO, [C\,{\sc ii}], [O\,{\sc iii}]) as well as continuum observations.

\item Planned \hi surveys (LADUMA, DINGO) on future telescopes will detect some ($\sim4\--20$) lenses, which in the case of LADUMA may require consideration when determining $\Omega_{\rm HI}(z\sim1)$. However, none of these surveys are particularly optimal for detecting lensed systems. Lensing does however enable the detection much lower \hi mass systems at high redshift, enhancing the scientific output of planned surveys.

\item Current telescopes (e.g. GMRT, VLA) fitted with their new broadband receivers, and most definitely future telescopes (e.g. MeerKAT), are very well placed to detect lensed \hi in targeted surveys and expand our view of cold neutral gas to higher redshift.

\item The depth and area of the SKA1-MID surveys will lead to efficient lens selection at high redshift. Overall, these surveys are predicted to detect over an order of magnitude more \hi lenses \mbox{($\sim0.8-2.2 \times 10^4$)} than the presently known lens population at any wavelength.

\item The large solid angles covered by \hi disks leads to significant differential magnification in some lensed sources. Appropriately optimized three-dimensional source finders may therefore enhance the detection rates presented here. 

\end{enumerate}

\section*{Acknowledgements}

We thank Masamune Oguri for making the {\tt glafic} software package publicly available and for providing very helpful advice. We also thank the anonymous referee for their comments. The financial assistance of the South African SKA Project (SKA SA) towards this research is hereby acknowledged. Opinions expressed and conclusions arrived at are those of the author and are not necessarily to be attributed to the SKA SA. www.ska.ac.za.

\end{document}